\documentclass [12pt]{article}
\usepackage{amsmath,amssymb}
\usepackage[dvips]{graphicx}
\RequirePackage[english, english]{babel}

\newcommand{\bs}{\begin{subequations}}
\newcommand{\es}{\end{subequations}}

\def \myfigures #1#2#3#4#5#6#7#8
{\begin{figure}[ht]
    \begin{center}
        \includegraphics[width=#2 \textwidth]{#1.eps}
        \hfill
        \includegraphics[width=#6 \textwidth]{#5.eps}
        \parbox[t]{#4\textwidth}{\caption {#3}\label{#1}}
        \hfill
        \parbox[t]{#8\textwidth}{\caption {#7}\label{#5}}
    \end{center}
\end{figure} }
\def \figr #1#2#3
{   \begin{figure}[ht]
        \centering\includegraphics[width=#2 \textwidth]{#1.eps}
        {\caption {#3}\label{#1}}
    \end{figure}
}   % #1 = имя файла, #2 = процент ширины страницы, #3 = caption.
\begin{document}
\title{Coordinate transformation in the model of long Josephson
junctions: geometrically equivalent Josephson junctions}

\author{E. G. Semerdzhieva\thanks{E-mail: elis@jinr.ru}\\[-1.mm]
\small\it Plovdiv University, Plovdiv 4000, Bulgaria, \\
%\small\it Joint Institute for Physical Research, Dubna 141980, Russia;\\\\
T. L. Boyadzhiev\thanks{E-mail: todorlb@jinr.ru} \\[-1.mm]
\small\it Joint Institute for Nuclear Research, Dubna 141980, Russia,\\
%\small\it Софийский университет, София, Болгария; \\\\
Yu. M. Shukrinov\thanks{E-mail: shukrinv@thsun1.jinr.ru}\\[-1.mm]
\small\it Joint Institute for Nuclear Research, Dubna 141980, Russia.}
%\small\it Physical Technical Institute, Dushanbe 734063,
%Tajikistan.}

\date{\small\it Submitted March 17, 2005\\ Fiz.\ Nizk.\ Temp.\ {\bf 31}, 1110--1116 (October 2005)}

\maketitle

\begin{abstract}

\pagenumbering{arabic}\thispagestyle{myheadings}

The transition from the model of a long Josephson junction of variable
width to the model of a junction with a coordinate-dependent Josephson
current amplitude is effected through a coordinate transformation. This
establishes the correspondence between the classes of Josephson junctions
of variable width and quasi-one-dimensio\-nal junctions with a variable
thickness of the barrier layer. It is shown that for a junction of
exponentially varying width the barrier layer of the equivalent
quasi-one-dimensional junction has a distributed resistive
inhomogeneity that acts as an attractor for magnetic flux vortices. The
curve of the critical current versus magnetic field for a Josephson
junction with a resistive microinhomogeneity is constructed with the
aid of a numerical simulation, and a comparison is made with the
critical curve of a junction of exponentially varying width. The
possibility of replacing a distributed inhomogeneity in a Josephson
junction by a local inhomogeneity at the end of the junction is thereby
demonstrated; this can have certain advantages from a technological
point of view.

\end{abstract}

\section*{1. INTRODUCTION}

An interesting research topic in the physics of Josephson junctions (JJs) and one that has received intensive development in recent years is the influence of a potential due to an external field or to the geometry on the motion of vortices in long junctions. As examples one can cite the study of the potential created by a spatial variation of the radial component of the magnetic field in an annular junction \cite{gls_91} and of a field-induced saw-tooth potential in the same annular junction \cite{c_01} and the study of the fluxon qubit in a modified annular (heart-shaped) geometry \cite{wklfu_00}. Potentials produced by spatial variation of the width of the junction, corresponding to a force acting on the fluxon in the direction of narrowing, were studied in Refs. \cite{cmc_02} -- \cite{ssb_05}. Unidirectional motion of fluxons has a certain advantage in the design of oscillators based on the motion of fluxons \cite{k_01,n_85,k_97}. In particular, an exponential variation of the junction width provides better impedance matching with an output load \cite{cmc_02} and allows one to avoid the chaotic regimes inherent to rectangular junctions. The structure and stability of static magnetic flux vortices and the corresponding critical curves in long JJs of exponentially varying width were investigated in Refs. \cite{cmc_02,bcs_96,sbs_04}, and \cite{ssb_05}.

In this paper it is shown that effects arising in a junction of variable width also take place for JJs whose barrier layer contains a resistive inhomogeneity. The critical curves --- critical current versus magnetic field --- are constructed numerically for long JJs of exponentially varying width and rectangular junctions with a localized resistive inhomogeneity. From a technological standpoint JJs with a resistive inhomogeneity can be preferable in some cases.

\section*{2. COORDINATE TRANSFORMATION}

We consider a long JJ whose dimension along the $y$ axis (width) is a smooth function $W(x)$ on the interval $x\in[0,L]$. Here $L$ is the length of the junction (all of the quantities used are rendered dimensionless in the standard way \cite{kkl_86}). We assume that
\[\max \mathop {W(x)}\limits_{x \in [0,L]}  \ll L\,.\]

We introduce the shape function $\sigma (x)$ of the junction with the aid of the relation
\begin{equation}\label{sigma}
    \sigma(x) = -\frac{W_x(x)}{W(x)}\,,
\end{equation}
Here and below a subscript is used to denote differentiation with respect to the correspon\-ding independent variable. The case $W_{x}(x)\equiv 0$ corresponds to a rectangular junction. Then the boundary-value problem for the static magnetic flux $\varphi (x)$ in such a JJ is written in the form
\bs\label{stat}
    \begin{gather}
       -\varphi_{xx} + \sigma(x)\,\varphi_x + \sin \varphi + g(x) = 0\,, \label{steq}\\
        \varphi_x(0) = h_0, \quad \varphi_x(L) = h_L\,.% + \varkappa_{l} l\,\gamma_B \,. \label{bcb}
    \end{gather}
\es
The concrete expressions for $q(x)$, $h_0$, and $h_{L}$ depend, in particular, on the means of injection of the external current $\gamma$. For junctions with an overlapping geometry, as are considered below, the base current $\gamma (x)$ flows through the whole junction. Consequently, $g(x) = \gamma(x) - \sigma h_{B}$, $h_0 = h_{B}$, $h_{L} = h_{B}$ ($h_{B}$ is the external magnetic field, which is directed along the $y$ axis in the plane of the barrier layer). The term $-\sigma h_{B}$ is due to the variable width of the junction. Neglecting surface effects, one can to a first approximation set $\gamma (x) = {\rm const}$. A derivation of Eq. \eqref{steq} in the general case is given in Ref. \cite{gsk_00}.

Let us consider an ordinary change of spatial coordinate:
\begin{equation}\label{tr}
    \xi = v(x)\,,
\end{equation}
where the variable $\xi \in [0,l]$, and the right-hand side is a
solution of the boundary-value problem
\bs\label{term}
    \begin{gather}
        -v_{xx} + \sigma(x)\, v_x  = 0\,,\\
        v(0) = 0,\quad \mathop{\min}\limits_{x \in [0,L]} v_x(x) = 1\,. \label{term2}
    \end{gather}
\es The parameters $L$ and $l$ are connected by the relation $l =
v(L)$. The first boundary condition in (4b) matches up the initial
points of the intervals of variation of the variables $x$ and $\xi
$, while the second serves for normalization (see below).

In the new independent variable the boundary value problem \eqref{stat} takes the form
\bs \label{steq1}
    \begin{gather}
        -\varphi_{\xi\xi} + j_D (\xi)\sin \varphi + j(\xi) = 0\,,\label{xieq}\\
        \varphi_\xi(0) = H_0\,, \quad \varphi_\xi(l) = H_L\,. \label{xibc}
    \end{gather}
\es Here $j_{D}(\xi )\equiv u_{\xi }^{2}(\xi )$, $j(\xi )\equiv
u_{\xi}^{2}(\xi)\,g(u(\xi ))$, $u(\xi )$ is the inverse function of $v(x)$, and also
    $$H_0 \equiv u_\xi(0) \,h_0, \quad H_L \equiv u_\xi(l)
    \,h_L\,.$$

From a formal standpoint Eq. \eqref{xieq} describes the magnetic flux distribution in a
one-dimensional JJ with a variable amplitude $j_{D}(\xi )$ of the Josephson current.
Such a junction is inhomogeneous \cite{galfil_84} --- the thickness of its barrier layer varies
along the length as $d(\xi )\sim u_{\xi}^{-2}(\xi)$. The term $j(\xi)$ is a current that
varies along the junction, generated by variation of the geometry, external magnetic
field, and external current.

Thus the transformation \eqref{tr}, \eqref{term} establishes a relation between a JJ of variable width and a ``one-dimensional'' junction having a thickness of the insulating layer that varies along the junction. For brevity we shall speak of $x$ junctions and $\xi$ junctions.

For junctions of exponentially diminishing width (referred to below as EJJs), which were considered in Refs.\ \cite{bcs_96} -- \cite{sbs_04}, we have $\sigma  = {\rm const} \geq 0$. Then $W(x) = W_0 \exp(-\sigma x)$, where $W_0$ is the width of the JJ at the $x = 0$ end. In this case the transformation (3) takes the form
$$\xi = \frac{1}{\sigma} \left(e^{\sigma x} - 1 \right)\,.$$
Then for the right-hand boundary of the $\xi $ junction we find $l = ({\rm e}^{\sigma
L}-1)/\sigma >L$, i.e., the $\xi$ junction is longer than the corresponding $x$ junction.

In the particular case considered, a change of the normalized amplitude of the Joseph\-son current in $\xi$-junction
\begin{equation}\label{jd}
    j_D(\xi)\equiv u_\xi^2(\xi)= \frac{1}{\left(1+\sigma\xi\right)^2}\,,
\end{equation}
can occur as a result of a barrier-layer thickness that varies smoothly along the junction. Here the amplitude is maximum (and the barrier-layer thickness minimum) at the left end of the junction (see the second condition in Eq. \eqref{term2}) and minimum (maximum, respectively) at the right end. Thus a resistive inhomogeneity is an attractor \cite{galfil_84} for magnetic flux
distributions in the junction, drawing the latter toward the right end, as has been observed experimentally \cite{cmc_02} and confirmed by numerical simulation \cite{cmc_02,sbs_04,ssb_05}.

\section*{3. MODELS OF A JUNCTION WITH A RESISTIVE INHOMOGENEITY}

It is of interest to investigate the possibility of replacing an inhomogeneity distributed over the whole length of the junction by an inhomogeneity localized near the right end. The simplest choice is a point inhomogeneity of ``strength'' $\mu  \geq 0$ at the right end of the junction. In that case \cite{galfil_84}
$$j_D(x) = 1 - \mu\, \delta(L-x)\,.$$
Here $\delta (x)$ is the Dirac delta function. A JJ with such an
inhomogeneity will be denoted below as $\delta $JJ. For $\mu  = 0$
the junction is homogeneous. The corresponding boundary-value
problem in the case of overlapping geometry is described in the
form [$x \in(0,L)]$
\bs \label{delta}
  \begin{gather}
    -\varphi_{xx} + \sin \varphi + \gamma  = 0, \label{deq}\\
    \varphi_x(0) = h_e ,\;\varphi_x(L) + \mu \sin\varphi(L) = h_e . \label{dbc}
  \end{gather}
\es We note that independently of the junction geometry the
presence of a point inhomoge\-neity at the right end leads to nonlinear boundary conditions.

Another example of an inhomogeneity that will be considered below is a narrow rectangular
well of width $\Delta  \ll L$ and depth $1-\kappa $:
$$j_D(x) = 1 - (1 - \kappa)\,\theta(x+\Delta-L)\,.$$
Here $\theta(x)$ is the Heaviside step function. The parameter $\kappa \geq 0$ specifies
the fraction of the Josephson current through the inhomogeneity. We note that the
inhomogeneity of the barrier layer represents a microresistor only for $\kappa <1$. In
particular, if $\kappa  = 0$ the Josephson current is absent on the segment $[L-\Delta
,L]$. The choice $\kappa  = 1$ means that the thickness of the barrier layer of the
junction is constant, and the amplitude $j_{D}(x) = 1$ for $x \in(0,L]$. For $\kappa
>1$ the inhomogeneity is a microshort that repels magnetic flux vortices \cite{galfil_84}.

The corresponding boundary-value problem for the magnetic flux
$\varphi (x)$ in the case of overlapping geometry is written in
the form
\bs \label{over}
 \begin{gather}
    -\varphi_{xx} + j_D (x)\sin \varphi - \gamma  = 0, \label{oeq}\\
    \varphi_x(0) = h_e ,\;\varphi_x(L) = h_e. \label{obc}
 \end{gather}
\es

In a numerical simulation of the magnetic-field dependence of the critical current, transitions from the superconducting to the resistive regime are interpreted mathemati\-cally as bifurcations of
the magnetic flux in the junction upon variation of the
parameters \cite{galfil_84}. For studying the stability, each solution
$\varphi (x)$ of the boundary-value problems \eqref{delta},
\eqref{over} is associated to a regular Sturm--Liouville (S--L)
problem with a potential that depends on the concrete solution
(see the details in Refs. \cite{galfil_84} -- \cite{tlb_02}). The S--L eigenvalue
problems allow one to assess the stability or instability of
$\varphi (x)$. The boundary conditions are determined by the
choice of model. In particular, for model \eqref{delta} with a
point inhomogeneity the S--L problem takes the form \bs
\label{dslp}
 \begin{gather}
   -\psi_{xx}  + q(x)\,\psi  = \lambda \,\psi\,,\label{dslpeq} \\
    \psi_x(0) = 0\,, \quad \psi_x(L) +\mu\,\cos\varphi(L)\,\psi(L) = 0\,,
 \end{gather}
\es where the potential is determined by the expression $q(x) =
\cos\varphi (x)$.

For model \eqref{over} with a rectangular inhomogeneity the S--L
problem is \bs \label{slp}
    \begin{gather}
        -\psi_{xx}  + q(x)\,\psi  = \lambda \psi\,,\label{slpeq} \\
        \psi_x(0) = 0\,, \quad \psi_x(L) = 0\,,
    \end{gather}
\es In the case under study the potential $q(x) =
j_{D}(x)\cos\varphi (x)$.

The wave function $\psi (x)$ must satisfy the normalization condition
\begin{equation}\label{norm}
    \int\limits_0^L \psi^2(x)\,d x - 1 = 0\,.
\end{equation}

We note that problems \eqref{dslp} and \eqref{slp} have a discrete
non-degenerate spectrum bounded from below. In addition, since the
solutions of the boundary-value problems \eqref{delta} and
\eqref{dslp} depend smoothly on the parameters $h_{B}$ and $\gamma
$, the eigenvalues and eigenfunctions of equations \eqref{dslp}
and \eqref{slp} also depend on these parameters. Let $\lambda
_{\rm min}(h_{B},\gamma )$ be the minimum eigenvalue corresponding
to a certain distribution $\varphi (x)$. The point with
coordinates $h_{B}$ and $\gamma $ on the $(h_{B},\gamma )$ plane
is called \cite{galfil_84} a bifurcation point of $\psi (x)$ if the
following condition holds:
\begin{equation}\label{crc}
    \lambda_{min}(\gamma, h_e) = 0\,.
\end{equation}

The bifurcation curve of any magnetic flux distribution in a junction is the geometric locus of bifurcation points of a concrete vortex, corresponding to pairs $(h_{B},\gamma )$ for which Eq. \eqref{crc} holds.

In view of the fact that for specified $h_{B}$ and $\gamma $ the boundary-value problems \eqref{delta} and \eqref{dslp}, as a rule, have more than one solution, the critical curve of the junction as a whole is constructed as the envelope of bifurcation curves for different solutions. In other words, the critical curve of the junction consists of pieces of the bifurcation curves corresponding to different distributions with the highest critical current $\gamma $ for the given field $h_{B}$.

For calculation of relation \eqref{crc} we used the method proposed in Ref. \cite{bpp_88}: the system of equations \eqref{delta} and \eqref{dslp} [respectively Eq. \eqref{over}, \eqref{slp}] at fixed $\lambda $ and field $h_{B}$ (and current $\gamma )$ is solved as a nonlinear eigenvalue problem with a spectral parameter $\gamma $ (respectively, $h_{B}$). This method of seeking bifurcation points and its generalization was used in Ref. \cite{tlb_02} for solving a wide class of physical problems.

\section*{4. NUMERICAL RESULTS}

Let us compare the results of numerical simulations for constructing the critical curves for the model of a JJ in an overlapping geometry with an exponentially varying width and a model with an inhomogeneity on the right end.
\myfigures{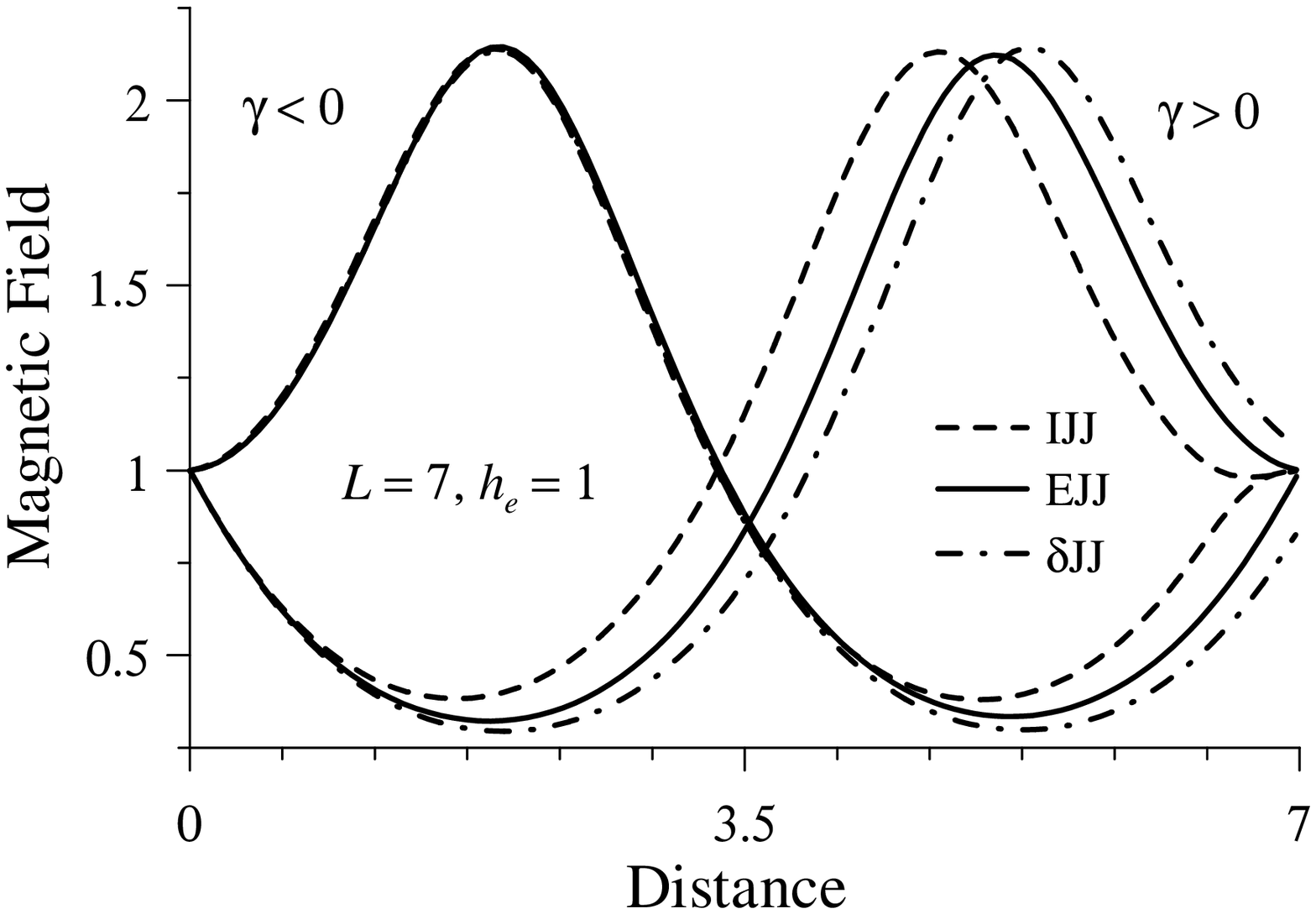}{0.48}{The magnetic field of the fluxon $\Phi^{1}$ for three models of the inhomogeneity.}{0.48}
{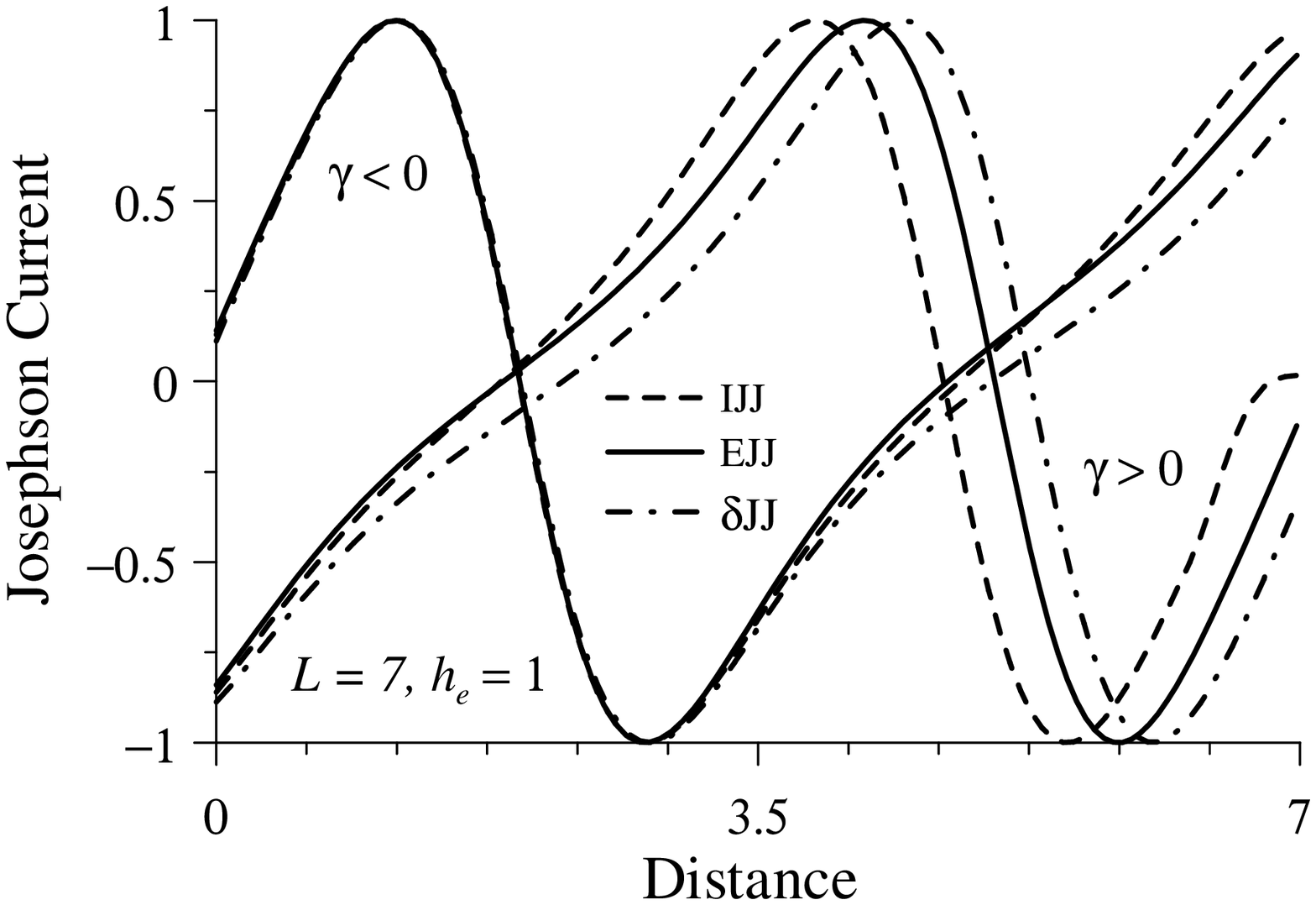}{0.48}{Distribution of the Josephson current for three models of the inhomogeneity.}{0.48}

Figure\ \ref{comp_mfield7} shows the ``bifurcation'' distributions of the magnetic field $\varphi_{x}(x)$ for the single fluxon $\Phi ^{1}$ along the junction for three JJ models (EJJ,
IJJ, and $\delta $JJ) for $\lambda_{\rm min} = 10^{-4}$, i.e., before the destruction of the fluxon by the current $\gamma $. The corresponding distributions of the Josephson current
$j_{D}(x)\sin\varphi $ in the JJ are presented in Fig.\ \ref{comp_jc7}.

We note the good qualitative agreement of the curves. Their quantitative difference is observed predominantly in the region where the amplitude of the Josephson current is minimum; this is due to the form of the inhomogeneity (attractor).

Figure\ \ref{exp_inh_o} demonstrates the critical curves of the form \eqref{crc} corresponding to stable solutions of the boundary-value problems \eqref{stat} and \eqref{over}. The solid curves correspond to distributions of the magnetic flux for the model of a JJ with a width that varies by an exponential law (EJJ) with a coefficient $\sigma = 0.07$, the dashed curves are for models of an inhomogeneous JJ with a finite rectangular inhomogeneity (IJJ) with parameters $\Delta  = 0.7$, $\kappa  = 0.1$. Each critical curve reflects a certain closed figure on the $(h_{B},\gamma )$ plane, the currents $h_{B} = h_{\rm min}$ and $h_{B} = h_{\rm max}$ are the lower and upper critical magnetic fields for the vortex under study. We note that because of the presence of an additional ``geometric'' current $\sigma (\varphi_{x} - h_{B})$ in the EJJ model the critical value of the current $\gamma $ corresponding to the two indicated values of $h_{B}$ is nonzero.
\figr{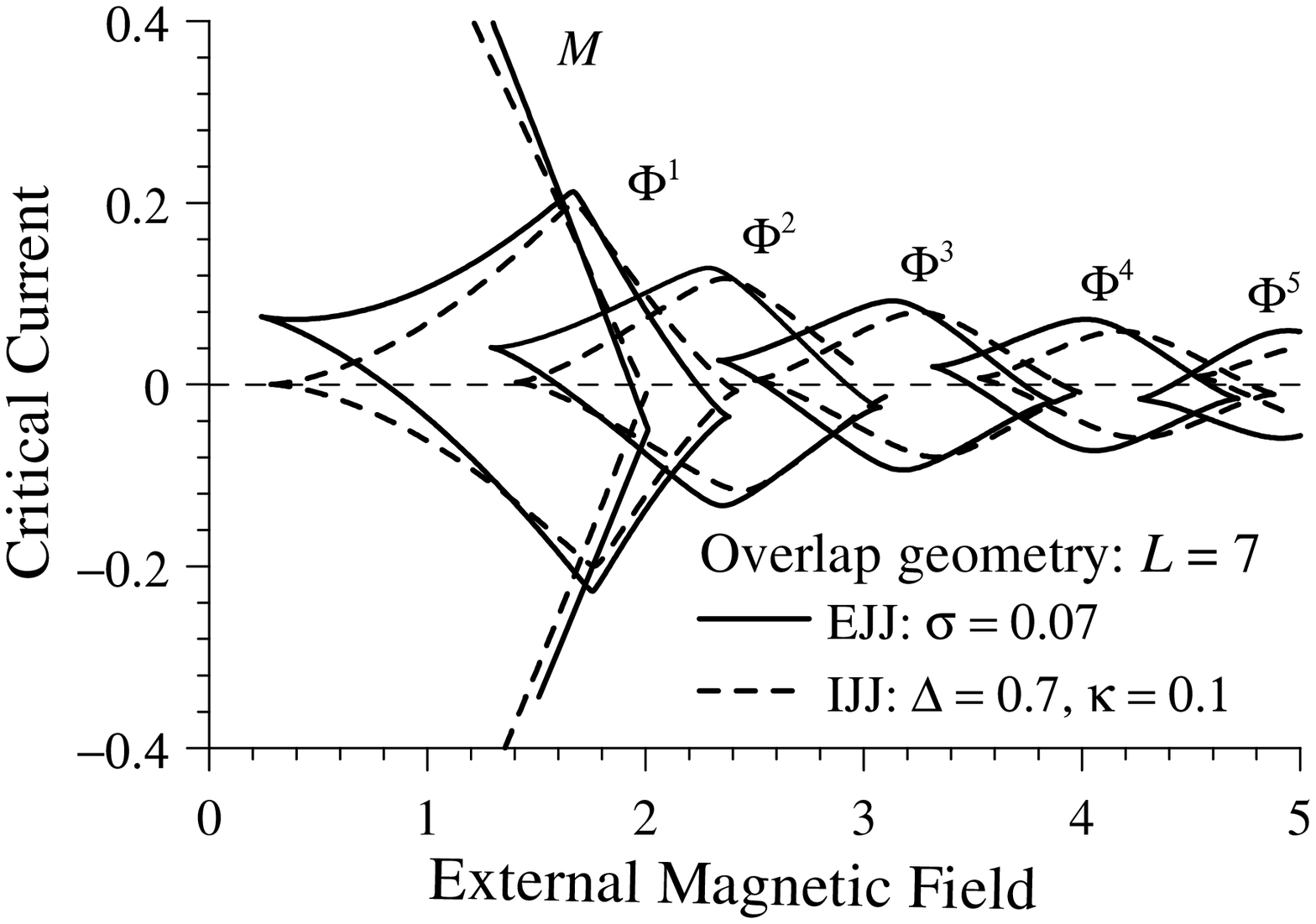}{0.5}{Bifurcation curves for a JJ of variable width and a JJ with a rectangular inhomogeneity.}

A numerical simulation shows that, unlike a JJ with a resistive inhomogeneity within the segment $(0,L)$ (Refs. \cite{bpp_88} and \cite{tlb_02}), in a junction with an inhomogeneity at the end, even for rather large values of $h_{B}$ (the solutions were checked numerically to $h_{B} = 10)$ there are no stable mixed fluxon--antifluxon vortices contributing to the critical curve.
Therefore the critical curves of the junction as a whole consist only of segments of the critical curves for a Meissner ($M$) distribution and ``pure'' $n$-fluxon vortices $\Phi ^{n}$, $n = 1,2,\ldots $. The maxima of the critical curves fall off monotonically with increasing field $h_{B}$ (in junctions with an internal inhomogeneity the maxima of the critical curves fall off non-monotonically because of the stabilizing influence \cite{tlb_02} of the inhomogeneity on the mixed fluxon--antifluxon vortex pairs).

Let us illustrate the process of vortex destruction (the transition to an unstable state) upon variation of the current $\gamma$ for the case of the IJJ model. In a field $h_{B} = 1$ at a current $\gamma \geq 0$, two stable distributions of magnetic flux can exist in the model: $M$ and $\Phi ^{1}$. As the current increases, the main fluxon $\Phi ^{1}$ is destroyed first (the critical current for it is $\gamma_{\rm cr}(\Phi ^{1})\approx 0.062$), and then the Meissner distribution. Thus for $h_{B} = 1$ the critical current of the junction $\gamma_{\rm cr} = \gamma_{\rm cr}({\rm M})\approx 0.597$. For $h_{B} = 1.9$ the boundary-value problem \eqref{over} has three stable solutions --- $M$, $\Phi ^{1}$, and $\Phi ^{2}$, which occur in the following succession with increasing $\gamma $: $\Phi ^{2}\rightarrow M \rightarrow \Phi ^{1}$. The critical current of the junction in this case is $\gamma_{\rm cr} = \gamma_{\rm cr}(\Phi^{1})\approx 0.157$. Analogously, for $h_{B} = 2.2$ we have $\gamma_{\rm cr} \gamma_{\rm cr}(\Phi ^{2})\approx 0.092$.

\figr{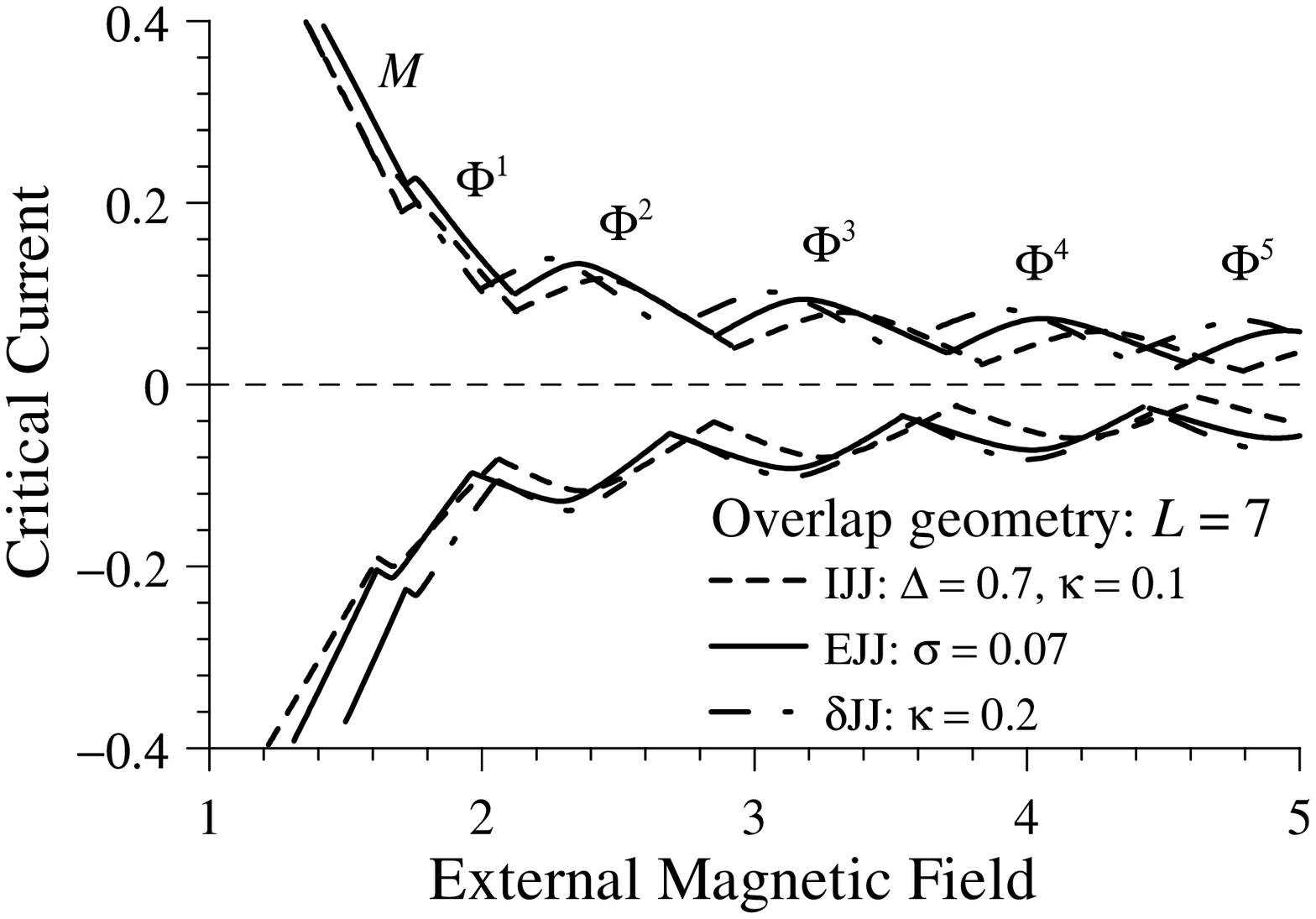}{0.5}{Critical curves of the current versus magnetic field for three model inhomogeneities. }
The critical curves of the current versus magnetic field for the three JJ models discussed above are exhibited in Fig.\ \ref{exp_inh_delta}. The dot-and-dash line corresponds to the model of a JJ with a point inhomogeneity ($\mu = 0.2)$ on the right end ($\delta $JJ). Note the good qualitative agreement of the critical curves for the three models. For values of $h_{B}$ that are not too high, the leftward displacement of the critical curves of the individual vortices by the ``geometric'' current in the EJJ model has only a weak influence on the behavior of the critical curve of the junction as a whole.
\myfigures{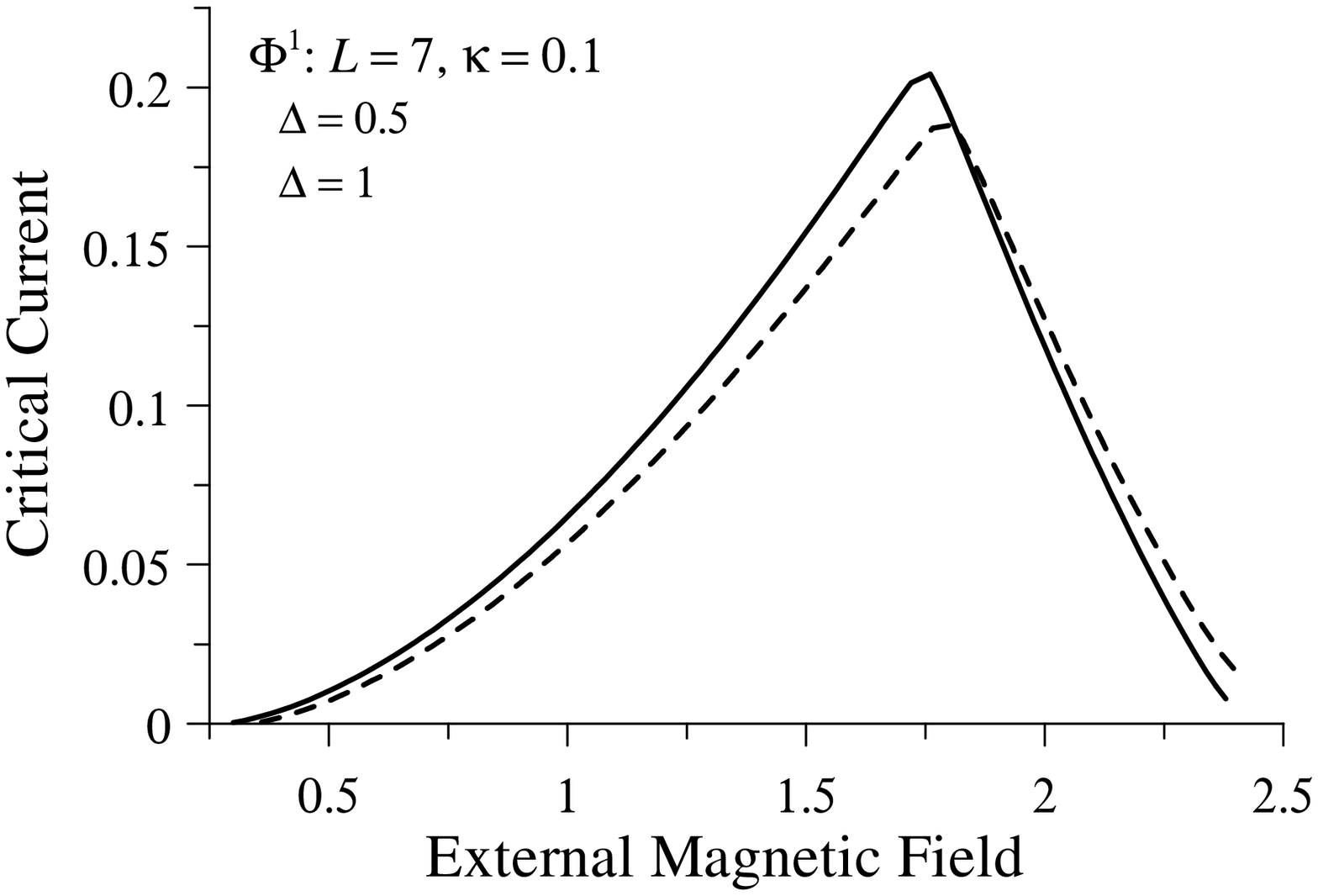}{0.48}{Influence of the parameter $\Delta $ on the bifurcation curve of $\Phi^{1}$.}{0.48}
{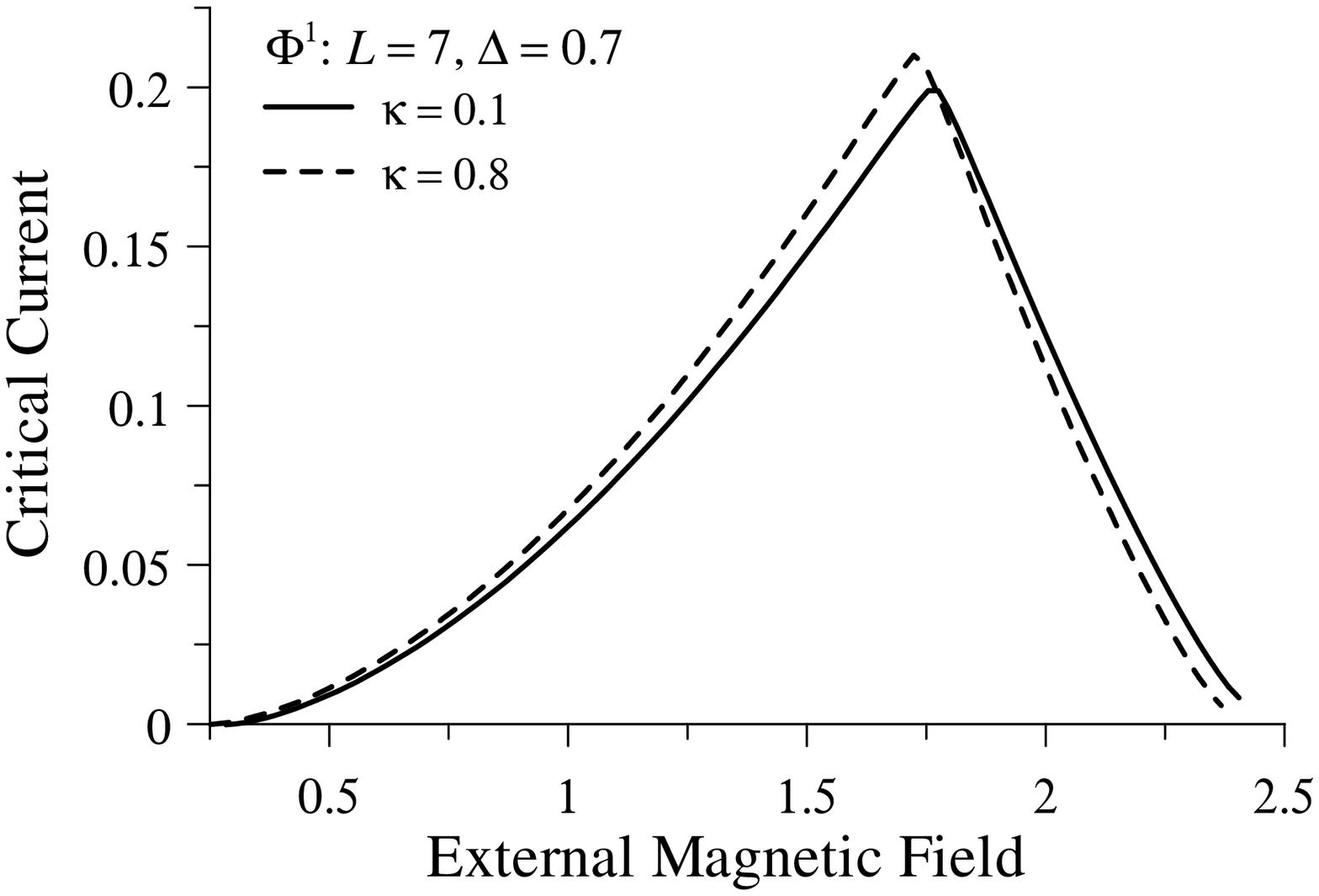}{0.48}{Influence of the parameter $\kappa$ on the bifurcation curve of $\Phi ^{1}$.}{0.48}

We note that variation of the parameters of a rectangular inhomogeneity (of width $\Delta $ and depth $\kappa )$ has a weak influence on the numerical results, i.e., the boundary-value problem \eqref{over} is structurally stable upon variation of those two parameters. This assertion is illustrated in Fig.\ \ref{comp_crc}, which shows the dependence \eqref{crc} for the main fluxon $\Phi^{1}$ at $\gamma_{\rm cr}\geq 0$, $\kappa  = 0.1$ and values of $\Delta = 0.5$ and $\Delta  = 1$. Analogously, Fig.\ \ref{comp_crc_ka} shows the dependence \eqref{crc} for the main fluxon $\Phi ^{1}$ for $\gamma_{\rm cr} \geq 0$, $\Delta  = 0.7$ and parameter values $\kappa  = 0.1$ and $\kappa  = 0.8$.

\section*{CONCLUSION}

In this paper we have shown that every JJ of variable width can be placed in correspon\-den\-ce with a quasi-one-dimensional junction with a variable thickness of the barrier layer. In the case of junctions whose width varies by an exponential law, the barrier layer of the corresponding quasi-one-dimensional JJ contains a resistive inhomogeneity distributed along the junction and acting as an attractor for magnetic flux vortices.

The curves of the critical current versus magnetic field obtained from the numerical simulation demonstrate that such an inhomogeneity can be replaced by a local inhomoge\-nei\-ty at the end of the JJ, which may have certain advantages from a technological point of view.

The authors thank Prof. \ Yu. Kolesnichenko (B. Verkin Institute for Low Temperature Physics and Engineering, Kharkov) for helpful discussions.

\noindent Translated by Steve Torstveit.

\end{document}